\documentclass[aps,prd,12point,twocolumn,nofootinbib,showpacs,superscriptaddress]{revtex4-2}

\usepackage{graphicx}
\usepackage{dcolumn}
\usepackage{tabu}
\usepackage{amsmath,amssymb}
\usepackage{float}
\usepackage{natbib}
\usepackage{balance}
\usepackage[normalem]{ulem}
\usepackage{braket}
\usepackage{upgreek}
\usepackage{mathrsfs}
\usepackage{bm}
\usepackage{lipsum}
\usepackage{comment}
\usepackage{xfrac}
\usepackage[dvipsnames]{xcolor}
\usepackage[title]{appendix}
\usepackage{txfonts}
\usepackage{xcolor}
\usepackage{slashed}
\usepackage[breaklinks=true,pdfborder={0 0 0},bookmarksopen]{hyperref}
\hypersetup{
	colorlinks   = true, 
	urlcolor     = magenta, 
	linkcolor    = blue, 
	citecolor   = Red 
}

\begin{document}

\title{Spin precession and neutrino helicity flip under the influence of a gravitational plane wave}

\author{R. Saadati}
\email{rsaadati@ubishops.ca}
\affiliation{Department of Physics \& Astronomy,
Bishop's University,\\
2600 College Street, Sherbrooke, QC, J1M 1Z7, Canada}

\author{F. Hammad}
\email{fhammad@ubishops.ca}
\affiliation{Department of Physics \& Astronomy,
Bishop's University,\\
2600 College Street, Sherbrooke, QC, J1M 1Z7, Canada}
\affiliation{Physics Department, Champlain College-Lennoxville,\\
2580 College Street, Sherbrooke, QC, J1M 0C8, Canada}

\begin{abstract}
We recently studied spin precession in various stationary and axisymmetric spacetimes and applied it to the case of neutrinos propagating in those spacetimes. In this paper, the study of spin precession is extended to the case of a spinning particle propagating within the spacetime of a time-dependent gravitational plane wave. First, the angular velocity of spin precession of a particle propagating in a geodesic along an arbitrary direction relative to the gravitational wave is derived. Our general result is then used to derive the helicity flip probability for neutrinos freely propagating along such an arbitrary direction within the gravitational wave.
\end{abstract}

\maketitle
\section{Introduction}\label{sec:Intro}
Neutrinos represent a potential probe for gravitational physics. Their weak interaction with matter and their nonzero mass make them one of the ideal messengers in astronomy and astrophysics. We have recently studied the effect of spin precession within various spacetimes \cite{Submitted}. We then made use of those results to derive the probability for a left-handed neutrino to flip its helicity and become a right-handed neutrino as it propagates within those spacetimes. That study was, nevertheless, restricted to stationary and axisymmetric spacetimes. In view of a broader application to modern multi-messenger astrophysics \cite{MMA}, however, the effect a gravitational wave would have on neutrinos spin flip is extremely important as well. In fact, it is well known that gravitational waves are always accompanied by bursts of neutrinos emitted from neutron star mergers as well as from black hole mergers \cite{LigoNeutrinos}. It is our aim in this paper to fill in this gap by working out the probability for left-handed neutrinos to become right-handed neutrinos as they propagate within the ripples of spacetime caused by the passage of a gravitational wave before they reach the detector. 

In Ref.\,\cite{Dvornikov2019}, Dvornikov has examined the problem of spin precession of neutrinos under the influence of a gravitational wave. That study was confined to the case of particles moving parallel to the gravitational wave. In this paper we make the most general study by allowing the particles to propagate in an arbitrary direction relative to the direction of propagation of the plane wave. The probability that the neutrinos flip their spin is then worked out. As neutrinos are always emitted by sources and detected by detectors in the form of left-handed particles, a helicity flip would entail a non-detection of the particles, which would then show up as a deficit in the total number of the expected neutrinos to be detected. Given that gravitational waves are expected to travel at the speed of light, whereas neutrinos are particles having a mass, a relative velocity of neutrinos with respect to those waves would be detected. Besides, the polarization of the waves should also be encoded in such a probability. For the sake of illustration and simplicity, however, we shall restrict ourselves here to linearly polarized gravitational waves. 

The rest of this paper is organized as follows. In Sec.\,\ref{Sec:PrecessionInGW}, we derive the general spin precession formula for a spinning particle moving in a geodesic along an arbitrary direction within the spacetime of a gravitational plane wave. In Sec.\,\ref{Sec:HelicityFlip}, we use that general formula to derive the probability for a left-handed neutrino to become a right-handed neutrino under the influence of the gravitational wave. In Sec.\,\ref{Sec:Conclusion}, we discuss our results in view of their application to observational astrophysics.

\section{Spin precession induced by a gravitational plane wave}\label{Sec:PrecessionInGW}
Spin precession in general relativity is described by the Mathisson-Papapetrou-Dixon (MPD) equations \cite{Mathisson,Papapetrou,Dixon1964,Dixon}. For point-like particles, like neutrinos, the so-called pole-dipole approximation of those equations is applicable as the size of the particle is much smaller than the characteristic scale of the gravitational field. As a consequence, the MPD equation describing the dynamics of the spin four-vector $S^{\mu}$ reduces to ${\rm D}S^\mu/{\rm d}\tau=0$ \cite{MTW}, where $\tau$ is the proper time of the particle and ${\rm D}$ is the covariant differentiation operator. 

To describe the absolute spin precession as measured relative to the fixed stars, one needs to cast the equation into the comoving frame of the neutrinos. To do so, one uses the spacetime comoving vierbeins $e_\mu^{\hat a}$, defined by $\eta_{{\hat a}{\hat b}}e_\mu^{\hat a}e_\nu^{\hat b}=g_{\mu\nu}$, where $g_{\mu\nu}$ is the curved spacetime metric and $\eta_{\mu\nu}$ is the Minkowski metric\footnote{The natural units for which $G=c=\hbar=1$ are used, and the metric signature $(-,+,+,+)$ is adopted. the Greek letters are used to denote spacetime indices, whereas the first Latin letters $(a,b)$ are used to denote tangent-space indices. The Latin letters $(i,j)$ are used to denote indices of the three-dimensional space. Hats over Latin letters denote indices in the comoving frame.}. The equation governing the spin three-vector $\textbf{S}$ (of spatial components $S^{\hat i}$) in the comoving frame of the particle reads
\begin{equation}\label{S^a Precession}
\frac{{\rm d}S^{\hat i}}{{\rm d}\tau}=-S_{\hat a}u^\mu\omega_\mu^{\,{\hat i}\hat{a}}.
\end{equation}
Here, $u^\mu$ is the four-velocity of the particle and $\omega_\mu^{\,{\hat a}{\hat b}}$ is the spin connection, defined in terms of the comoving vierbeins, their inverses $e_{\hat a}^\mu$ (such that $e^{\hat a}_\mu e^\mu_{\hat b}=\delta^{\hat a}_{\hat b}$, where $\delta_{\hat b}^{\hat a}$ is the Kronecker delta symbol), as well as the Christoffel symbols $\Gamma_{\mu\nu}^\lambda$ corresponding to the spacetime metric, by 
\begin{equation}\label{SpinConnectionDef}
\omega_\mu^{\,\,{\hat a}{\hat b}}=-e^{\nu{\hat  b}}\partial_\mu e^{\hat a}_\nu+e^{\nu {\hat b}}\Gamma_{\mu\nu}^\lambda e^{{\hat a}}_\lambda.
\end{equation}
The angular velocity three-vector $\boldsymbol{\Omega}$ of the spin precession is then extracted from Eq.\,(\ref{S^a Precession}) which is of the form ${\rm d}{\boldsymbol{S}}/{\rm d}\tau={\boldsymbol{\Omega}}\times{\boldsymbol{S}}$. The explicit components of $\boldsymbol{\Omega}$ thus read 
\begin{equation}\label{OmegaVector}
    \Omega_{\hat i}=\tfrac{1}{2}\varepsilon_{{\hat i}{\hat j}{\hat k}}u^\mu\omega_{\mu}^{\,{\hat j}{\hat k}},
\end{equation}
where $\varepsilon_{{\hat i}{\hat j}{\hat k}}$ stands for the totally antisymmetric Levi-Civita symbol. 

We shall apply this formula to particles moving along a geodesic in the spacetime of a gravitational plane wave propagating, for definiteness, along the $x$-direction. For that purpose, we adopt the following spacetime metric in the transverse-traceless gauge in the coordinates $(t,x,y,z)$ (see, e.g., \cite{Hartle}):
\begin{equation}\label{GW}
{\rm d}s^2=-{\rm d}t^2+{\rm d}x^2+\left(1-h_{+}\right){\rm d}y^2+\left(1+h_{+}\right){\rm d}z^2-2h_{\times}{\rm d}y\,{\rm d}z,
\end{equation}
where, for the wave chosen here to be elliptically polarized for generality, we have 
\begin{equation}
h_{+}=a\sin\left[k(t-x)+\theta\right],\qquad h_{\times}=b\sin\left[k(t-x)+\delta\right].
\end{equation}

For the general case, one needs to keep both the real amplitudes $a$ and $b$ ---\,as well as the real phases $\theta$ and $\delta$ that determine completely the state of polarization of the wave\,--- all arbitrary. The wave vector of the wave is denoted by $k$. The spacetime metric (\ref{GW}) has three Killing vector fields: the two spacelike Killing vector fields $K^2=(0,0,1,0)$ and $K^3=(0,0,0,1)$ along the $y$ and $z$ directions respectively, and the null Killing vector field $K^\chi=(1,1,0,0)$ along the advanced null direction $\chi=t+x$. The two spacelike Killing vectors $K^2$ and $K^3$ yield the conserved momenta per unit mass of the particle, $\alpha=g_{y\mu} u^\mu$ and $\beta=g_{z\mu} u^\mu$, respectively. The null Killing vector $K^\chi$ yields the conserved positive quantity $E=-(g_{0\mu} u^\mu+ g_{x\mu} u^\mu)$ that represents the difference between the energy per unit mass of the particle and the momentum per unit mass of the particle along the $x$ direction \cite{Felice}. Together with the normalization $g_{\mu\nu}u^\mu u^\nu=-1$ of the four-velocity, these conserved quantities lead to the following four-velocity vector of the particle along the geodesic of the spacetime \cite{Felice}:
\begin{align}\label{General4Velocity}
u^t&=\frac{1+E^2}{2E}+\frac{\alpha^2\left(1+h_+\right)+\beta^2\left(1-h_+\right)+2\alpha\beta h_{\times}}{2E\left(1-h_+^2-h_{\times}^2\right)},\nonumber\\
u^x&=\frac{1-E^2}{2E}+\frac{\alpha^2\left(1+h_+\right)+\beta^2\left(1-h_+\right)+2\alpha\beta h_{\times}}{2E\left(1-h_+^2-h_{\times}^2\right)},\nonumber\\
u^y&=\frac{\alpha\left(1+h_+\right)+\beta h_{\times}}{\left(1-h_+^2-h_{\times}^2\right)},\nonumber\\
u^z&=\frac{\beta\left(1-h_+\right)+\alpha h_{\times}}{\left(1-h_+^2-h_{\times}^2\right)}.
\end{align}
What determines whether the particle is moving in the same direction as the wave or in the opposite direction of the wave is the sign of component $u^x$, which itself depends on whether $E<1$ or $E>1$, respectively. This can be seen by setting $\alpha=\beta=0$ in Eq.\,(\ref{General4Velocity}) ---\,which corresponds to a particle moving parallel or antiparallel to the wave's direction of propagation\,--- leading to $u^x=(1-E^2)/2E$ and $u^y=u^z=0$. 

Using the metric components (\ref{GW}) and the components of the four-velocity as given in Eq.\,(\ref{General4Velocity}), we build the following four comoving vierbeins:
\begin{align}\label{Vierbeins}
    e_{\hat 0}^\mu&=\left(u^t,u^x,u^y,u^z\right),\nonumber\\
e_{\hat 1}^\mu&=\frac{1}{X}\left(u^x,u^t,0,0\right),\nonumber\\
e_{\hat2}^\mu&=\frac{1}{XY}\left(\beta u^t,\beta u^x,0,X^2 
  \right),\nonumber\\
  e_{\hat 3}^\mu&=\frac{\sqrt{-g}}{Y}\left(u^tu^y,u^xu^y,-\frac{Y^2}{g},u^z u^y-\frac{h_\times}{g}\right),
\end{align}
where $g$ is the determinant of the metric (\ref{GW}), and
\begin{align}
    X&=\sqrt{1+\alpha u^y+\beta u^z},\nonumber\\
    Y&=\sqrt{1+h_+-g(u^y)^2}{.}
\end{align}
The inverse vierbeins $e^{\hat a}_\mu$ are then found to be
\begin{align}\label{InverseVierbeins}
    e^{\hat 0}_\mu&=\left(u^t,-u^x,-\alpha,-\beta\right),\nonumber\\
    e^{\hat 1}_\mu&=\frac{1}{X}\left(-u^x,u^t,0,0\right),\nonumber\\
    e^{\hat 2}_\mu&=\frac{1}{XY}\left(-\beta u^t,\beta u^x,-X^2 h_{\times},X^2 \left(1+h_+\right)\right),\nonumber\\ e^{\hat 3}_\mu&=\frac{\sqrt{-g}u^y}{Y}\left(-u^t,u^x,\frac{1}{u^y}\!+\!(1\!-\!h_+)u^y\!-\!h_\times u^z,(1\!+\!h_+)u^z\!-\!h_\times u^y\right).
\end{align}
Using the explicit expressions of these vierbeins  and their inverses, as well as the Christoffel symbols (\ref{AppChristoffels}) computed in Appendix \ref{Sec:AppChristoffel}, we find that the particle's spin would precess with an angular velocity vector $\boldsymbol\Omega$ that has the components given below. No approximation is made in the computation that led to those long expressions. As such, these components of the angular velocity are exact and are valid for an arbitrary geodesic propagation of the spinning particle within the gravitational wave background. However, given the complexity of those expressions, some approximations will be made in the next section where these components of the angular velocity are used to derive the spin-flip probability for neutrinos propagating within the spacetime metric (\ref{GW}).
\begin{align}\label{Omega}
    \Omega_{\hat{1}}&=\frac{1}{2\sqrt{-g}XY^2}\Bigg\{YX^2\left(u^xh'_++u^t\dot{h}_\times\right)\nonumber\\
    &\quad+\left(u^xh'_+-u^t\dot{h}_\times\right)\left[g(u^y)^2(X^2-\beta u^z)+\beta(h_\times u^y+u^zY^2)\right]\nonumber\\
    &\quad-\beta\left(u^t\dot{h}_++u^xh'_+\right)\left[g u^y(u^z)^2+u^y Y^2-u^z h_\times\right]\nonumber\\
    &\quad-h_\times X^2\left(u^xh'_+-u^t\dot{h}_+\right)+2g u^xu^y u^z X^2h'_+\nonumber\\
    &\quad+2g u^t u^yX\left(X\dot{X}-\frac{X\dot{Y}}{Y}\right)\left[h_\times u^y-(1+h_+)u^z\right]\nonumber\\
    &\quad-2g\beta u^tu^yX\left(X\dot{X}+\frac{X\dot{Y}}{Y}\right)+2g\beta u^t u^y \left(u^t\dot{u}^t-u^x\dot{u}^x\right)\Bigg\},\nonumber\\
    \Omega_{\hat{2}}&=\frac{1}{2\sqrt{-g}XY}\Bigg[2gu^y(u^t)^2\dot{u}^x-2gu^xu^yu^t\dot{u}^t\nonumber\\
    &\quad+\left(u^t h'_++u^x\dot{h}_\times\right)\left(g(u^y)^2u^z-u^zY^2-h_\times u^y\right)\nonumber\\
    &\quad-\left(u^th'_++u^x\dot{h}_+\right)\left(g(u^z)^2u^y+u^yY^2-h_\times u^z\right),\nonumber\\
    \Omega_{\hat{3}}&=\frac{1}{2\sqrt{-g}XY^2}\Bigg\{YX^2\left(u^t\dot{h}_\times-u^xh'_\times\right)\nonumber\\
&\quad+\left(u^t\dot{h}_\times+u^xh'_\times\right)\left[g(u^y)^2(X^2-\beta u^z)+\beta(h_\times u^y+u^z Y^2)\right]\nonumber\\
&\quad+\beta\left(u^t\dot{h}_++u^xh'_+\right)\left[gu^y(u^z)^2+u^yY^2-u^zh_\times\right]\nonumber\\
&\quad-h_\times X^2\left(u^xh'_+-u^t\dot{h}_+\right)-2gu^xu^yu^zX^2h'_+\nonumber\\
&\quad+2gu^tu^yX\left(X\dot{X}+\frac{X\dot{Y}}{Y}\right)\left[h_\times u^y-(1+h_+)u^z\right]\nonumber\\
&\quad-2g\beta u^t u^y\left(X\dot{X}-\frac{X\dot{Y}}{Y}\right)+2g\beta u^tu^y\left(y^t\dot{u}^t-u^x\dot{u}^x\right)\Bigg\}.
\end{align}
\section{Neutrino helicity flip inside a gravitational plane wave}\label{Sec:HelicityFlip}
The probability for neutrinos helicity flip can be obtained using the effective Hamiltonian $H_{\rm eff}(\textbf{r})$ that arises from the spin-gravity interaction. The effective Hamiltonian $H_{\rm eff}(\textbf{r})$ is, in turn, obtained from the three Pauli matrices $\boldsymbol\sigma=(\sigma_1,\sigma_2,\sigma_3)$ and the spin precession angular velocity $\boldsymbol\Omega$ as, $H_{\rm eff}(\textbf{r})=\frac{1}{2}{\boldsymbol\sigma}.{\boldsymbol\Omega}$. For an initial spin state $\ket{S_{\rm in}}$, the probability for a neutrino to be found in the final spin state $\ket{S_{\rm fi}}$ is \cite{SO}
\begin{equation}\label{GeneralProbability}
\mathcal{P}(S_{\rm in}\rightarrow S_{\rm fi})=\left|\bra{S_{\rm fi}}{\textbf{T}}\exp\left[-i\int_{\tau_i}^{\tau_f}H_{\rm eff}({\textbf{r}}){\rm d}\tau\right]\ket{S_{\rm in}}\right|^2,
\end{equation}
where $\textbf{T}$ is the usual time-ordering operator and $\tau_i$ and $\tau_f$ are the initial and final proper times. For an initially left-handed neutrino moving with four-velocity $u^\mu=u^t(1,\vec{\varv})$, where $\vec{\varv}$ is the three-velocity of magnitude $\varv$, the normalized initial spin state can be written as $\ket{S_{\rm in}}=\frac{1}{\sqrt{2}}(\eta,-\xi e^{i\phi})$,
where $\eta=(1-\varv_z/\varv)^{1/2}$, $\xi=(1+\varv_z/\varv)^{1/2}$ and $e^{i\phi}=(\varv_x+i\varv_y)/(\varv\eta\xi)$. The final right-handed state of the neutrino then reads $\ket{S_{\rm fi}}=\frac{1}{\sqrt{2}}(\xi,\eta e^{i\phi})$.
For these initial and final spin states, formula (\ref{GeneralProbability}) yields the following probability for a left-handed neutrino $\ket{\nu_L}$ to transform into a right-handed neutrino $\ket{\nu_R}$:
\begin{align}\label{ExplicitProbabilityFormula}
&\mathcal{P}(\ket{\nu_L}\rightarrow \ket{\nu_R})=\left[\int_{\tau_i}^{\tau_f}\,\Omega\,{\rm d}\tau\right]^{-2}\sin^2\left[\tfrac{1}{2}\int_{\tau_i}^{\tau_f}\Omega\,{\rm d}\tau\right]\nonumber\\
&\quad\quad\times\left\{\left(\int_{\tau_i}^{\tau_f}\left[\eta\xi\Omega_z+\tfrac{1}{2}(\eta^2-\xi^2)\left(\Omega_x\cos\phi+\Omega_y\sin\phi\right)\right]{\rm d}\tau\right)^{2}\right.\nonumber\\
&\quad\quad\left.+\left[\int_{\tau_i}^{\tau_f}\tfrac{1}{2}\left(\eta^2+\xi^2\right)\left(\Omega_y\cos\phi-\Omega_x\sin\phi\right){\rm d}\tau\right]^2\right\}.
\end{align}
Here, $\Omega$ denotes the magnitude of the precession angular velocity vector. We shall now apply this formula to extract the helicity flip probability for neutrinos propagating within the gravitational wave (\ref{GW}).

To be able to easily perform the integration in Eq.\,(\ref{ExplicitProbabilityFormula}), we need to convert the integration over the proper time $\tau$ into an integration over the coordinate time $t$. For that purpose, we make use of the $t$-component of the four-velocity in Eq.\,(\ref{General4Velocity}). On the other hand, to deal with the presence of $x(t)$, we start by extracting such a function of time. For simplicity, but without a big loss of generality, we set in what follows $a=b$ and $\theta=\delta=0$ in the gravitational wave metric (\ref{GW}) so that $h_+=h_\times=a\sin[k(t-x)]$. 
The path of the particle along the $x$-axis we obtain from Eq.\,(\ref{General4Velocity}) in this case reads
\begin{equation}\label{dx/dtBeta=0}
\frac{{\rm d}x}{{\rm d}t}\!=\!\frac{\left(1-E^2\right)\left(1\!-\!h_+^2\!-\!h_{\times}^2\right)+\alpha^2\left(1\!+\!h_+\right)+\beta^2\left(1\!-\!h_+\right)+2\alpha\beta h_{\times}}{\left(1+E^2\right)\left(1\!-\!h_+^2\!-\!h_{\times}^2\right)+\alpha^2\left(1\!+\!h_+\right)+\beta^2\left(1\!-\!h_+\right)+2\alpha\beta h_{\times}}.
\end{equation}
The exact solution to this nonlinear first-order differential equation is:
\begin{align}\label{x(t)}
    x(t)&=\frac{A+(1-E^2)t}{1+E^2}- c_+ \tan^{-1}\left(\frac{\sqrt{2} a-\tan \left[\frac{1}{2} k (t-x\right]}{\sqrt{1-2 a^2}}\right)\nonumber\\
    &\quad-c_- \tan^{-1}\left(\frac{\sqrt{2} a+\tan \left[\frac{1}{2} k (t-x)\right]}{\sqrt{1-2 a^2}}\right),
\end{align}
where 
\begin{equation}
    c_{\pm}=\frac{\left(\sqrt{2}\pm2\right) \alpha ^2+2 \sqrt{2} \alpha  \beta -\left(\sqrt{2}\mp2\right) \beta ^2}{2 \sqrt{1-2 a^2}k \left(1+E^2\right)}.
\end{equation}
The constant of integration $A$ can be chosen such that $x=0$ at $t=0$. Implementing this condition leads, according to \eqref{x(t)}, to
\begin{equation}
    A=\frac{\sqrt{2}\left(\alpha ^2+2 \alpha  \beta -\beta ^2\right)}{\sqrt{1-2a^2} k} \tan ^{-1}\left(\frac{\sqrt{2} a}{\sqrt{1-2 a^2}}\right).
\end{equation}

Although isolating $x(t)$ from equation \eqref{x(t)} is not possible, an approximate solution can be found. In fact, for high-energy neutrinos, we have $E\ll1$ when they propagate with the wave and $E\gg1$ when they propagate in the opposite direction of the wave. But on the other hand, for the typical gravitational waves detected so far, we often have $a\sim10^{-(16\sim22)}$ and $k=2\pi/\lambda\sim10^{-14}$m$^{-1}$ \cite{GWReview1,GWReview2}. Therefore, we can expand the exact solution (\ref{x(t)}) for $a\ll1$ and isolate an approximate expression for $x(t)$ as a function of $t$. Keeping only terms that are at most of first order in $a$ in Eq.\,(\ref{x(t)}) allows us to isolate $x(t)$ in the form:
\begin{align}\label{FirstOrderx(t)}
    x(t)&=\left(\frac{1-E^2+\alpha ^2+\beta ^2}{1+E^2+\alpha ^2+\beta ^2}\right)t\nonumber\\
    &\quad+\frac{2a(\alpha ^2+2 \alpha  \beta -\beta ^2)}{k(1+E^2+\alpha ^2+\beta ^2)}\sin ^2\left(\frac{E^2 k t}{1\!+\!E^2\!+\!\alpha ^2\!+\!\beta ^2}\right)+\mathcal{O}(a^2).
\end{align}
This shows that the motion of the neutrinos along the $x$-axis is a uniform one, but slightly perturbed by an extremely tiny oscillatory motion of amplitude proportional to $a/k$. This expression of $x$ in terms of $t$ allows us to cast all the variables in the integrals in Eq.\,(\ref{ExplicitProbabilityFormula}) into functions of the single variable $t$ over which we must integrate. 

Going back to the integral in Eq.\,(\ref{ExplicitProbabilityFormula}) and substituting the expressions of $\eta$ and $\xi$ in terms of the speed $\varv$ and the velocity components $\varv_x,\varv_y$ and $\varv_z$, the integral takes the following compact form after replacing  ${\rm d}\tau$ by ${\rm d}t/u^t$:
\begin{align}\label{DetailedProbabilityFormula}
&\mathcal{P}(\ket{\nu_L}\rightarrow \ket{\nu_R})=\left[\int_{0}^{t}\frac{\Omega}{u^t}\,{\rm d}t\right]^{-2}\sin^2\left[\tfrac{1}{2}\int_{0}^{t}\frac{\Omega}{u^t}\,{\rm d}t\right]\nonumber\\
&\quad\qquad\qquad\times\left\{\left(\int_{0}^{t}\left[\sqrt{1-\frac{\varv_z^2}{\varv^2}}\frac{\Omega_z}{u^t}-\frac{\varv_z \left(\varv_x \Omega _x+\varv_y \Omega _y\right)}{\varv u^{t} \sqrt{\varv_x^2+\varv_y^2}}\right]{\rm d}t\right)^{2}\right.\nonumber\\
&\quad\qquad\qquad\quad\left.+\left[\int_{0}^{t}\,\frac{\varv_x \Omega _y-\varv_y \Omega _x}{u^{t}\sqrt{\varv_x^2+\varv_y^2}}{\rm d}t\right]^2\right\}.
\end{align}
We have set the limits of integration to be the initial time $t=0$, when the neutrino enters the wave, and an arbitrary later time $t$. The latter should be a time for which the neutrino is still in contact with the wave. Now, although an analytic evaluation of this integral using the full expressions of the components (\ref{Omega}) of the angular velocity vector $\boldsymbol\Omega$ might be performed numerically, the tremendous complexity of the expression that would result would merely prohibit us from easily accessing the physical content of the final answer. For this reason, we numerically evaluate the integral based on the first-order approximation (\ref{FirstOrderx(t)}) of the trajectory of the neutrinos. The result, written up to the second order in $a$, reads 
\begin{equation}\label{probability}
    \mathcal{P}(\ket{\nu_L}\rightarrow \ket{\nu_R})=
    \frac{a^2\left(c_1^2+c_2^2\right)}{16 E^4} \sin ^2\left(\frac{2 E^2 k t}{1+E^2+\alpha ^2+\beta ^2}\right),
\end{equation}
where $c_1$ and $c_2$ are constants whose full explicit expressions are given in Eqs.\,\eqref{c1} and \eqref{c2}, respectively. This is a very general result as it gives the probability at the second order in the amplitude $a$ of the gravitational wave for any type of propagation of neutrinos relative to the gravitational wave. 

The first remarkable thing we notice from this result is the oscillatory nature of the spin-flip probability. The period of the probability oscillation is always longer than that of the gravitational wave when the neutrinos are not propagating in the opposite direction of the wave. 
When high-energy neutrinos are propagating in the same direction as the wave, for which case we have $E\ll1$, the probability ceases to be oscillatory during the time interval in which the neutrinos are still inside the wave and tends to increase in that time like the second power of $t$. On the other hand, when high-energy neutrinos are propagating in the opposite direction of the wave, for which case we have $E\gg1$, the period of the probability oscillation tends to be half that of the gravitational wave that caused the spin precession. For low-energy neutrinos, the probability tends to increase like the second power of the time during which the neutrinos are inside the gravitational wave regardless of whether the neutrinos are propagating in the same or in the opposite direction of the wave.
\section{Summary and conclusion}\label{Sec:Conclusion}
We have studied spin precession induced by a gravitational plane wave on spinning particles freely propagating in a geodesic along an arbitrary direction relative to the gravitational wave. We computed the components of the general angular velocity vector of the precession before applying our result to the case of left-handed neutrinos. We derived the probability for the latter to become right-handed as they propagate within the wave. Given the complexity of the general angular velocity obtained for spin precession, we restricted our study to the special case of linearly polarized gravitational waves. The small amplitude of the latter allowed us to derive a probability for neutrinos to flip their spin. We found that such a probability is oscillatory and proportional to the square of the wave's amplitude. In addition, we found that the period of the probability oscillation varies depending on whether the neutrinos are propagating in the same or in the opposite direction of the wave. Furthermore, for low-energy neutrinos ---\,regardless of their direction of propagation\,--- and for high-energy neutrinos propagating in the same direction as the gravitational wave, we found that in both cases the probability ceases to be periodic during the time interval in which the neutrinos are inside the wave and increases instead as the second power of time.

The interesting fact about the spin-flip probability (\ref{probability}) is that it carries information not only about the kinematics and the direction of motion of the neutrinos relative to the gravitational wave, but it also carries information about the wavelength and the amplitude of the wave. This might be of particular importance for observational modern multi-messenger astrophysics, a study which will be conducted in more detail elsewhere.

For realistic gravitational waves, of course, given the many known potential astrophysical sources of such waves --- as well as all the possible departures from general relativity one might give allowance for \cite{Callister:2017ocg}, we do not generally expect to have $a = b$ nor $\theta = \delta = 0$. Indeed, stochastic gravitational-wave backgrounds arising from the superposition of many individually unresolvable gravitational-wave signals \cite{Callister:2017ocg} and gravitational waves from binary mergers of compact objects typically have different amplitudes and phases for the various polarization states. The components of the latter are often asymmetric, depending also on the orientation and inclination of the source \cite{LigoNeutrinos,GWReview2}. In this study, we assumed $a = b$ and $\theta = \delta = 0$ solely to simplify the analysis and be able to extract the main preliminary physical consequences from our integral \eqref{DetailedProbabilityFormula}.

Relaxing these simplifications would result in a more complex spin-flip probability expression, which would depend on the polarization states of the wave as well as on the relative phases $\theta$ and $\delta$ of the latter. Thus, the time dependence of the probability expression \eqref{DetailedProbabilityFormula} would be greatly modified in a way that reflects the asymmetry between the polarization modes of the gravitational wave. We expect that this more general treatment could indeed lead to additional oscillatory terms in Eq.\,\eqref{DetailedProbabilityFormula}, coming from Eq.\,\eqref{Omega} and from Eq.\,\eqref{dx/dtBeta=0} as follows. First, when fully taken into account in Eq.\,\eqref{dx/dtBeta=0}, $a \neq b$ and $\theta \neq \delta \neq 0$ lead to additional nontrivial oscillations of the neutrino’s motion along its trajectory. This has the effect, according to Eq.\,\eqref{DetailedProbabilityFormula}, of modulating the spin oscillations’ integrated response time inside the wave. Second, Eq.\,\eqref{Omega} introduces ---already at the level of the angular velocity of spin oscillation--- a coupling between, on the one hand, the velocity components of the neutrino, and on the other hand, the amplitudes $a$ and $b$ of the polarization states as well as the phases $\theta$ and $\delta$ of the latter. The unapproximated probability expression \eqref{DetailedProbabilityFormula} would thus display a complex dependence on the energy and direction of propagation of the neutrinos relative to the gravitational wave and on the polarization states of the latter even at the orders $\mathcal{O}(a^2)$ and $\mathcal{O}(b^2)$ considered here. These aspects centered around the effect of the gravitational wave polarization on neutrino spin-flip probabilities will also be the subject of future studies.

\section*{Acknowledgments}
This work was supported by the Natural Sciences and Engineering Research Council of Canada (NSERC) Discovery Grant No. RGPIN-2017-05388; and by the Fonds de Recherche du Québec - Nature et Technologies (FRQNT).

\appendix
\section{Christoffel symbols from the metric (\ref{GW}) }\label{Sec:AppChristoffel}The nonzero Christoffel symbols we extract from the metric components (\ref{GW}) are the following:
\begin{align}\label{AppChristoffels}
\Gamma_{02}^2&=-\frac{h_{\times}\dot{h}_{\times}+\left(1+h_+\right)\dot{h}_+}{2\left(1-h_+^2-h_{\times}^2\right)},\quad\quad \Gamma_{12}^2=-\frac{h_{\times}h'_{\times}+\left(1+h_+\right)h'_+}{2\left(1-h_+^2-h_{\times}^2\right)},\nonumber\\
\Gamma_{02}^3&=-\frac{h_{\times}\dot{h}_{+}+\left(1-h_+\right)\dot{h}_{\times}}{2\left(1-h_+^2-h_{\times}^2\right)},\quad\quad \Gamma_{12}^3=-\frac{h_{\times}h'_{+}+\left(1-h_+\right)h'_{\times}}{2\left(1-h_+^2-h_{\times}^2\right)},\nonumber\\
\Gamma_{03}^2&=\frac{h_{\times}\dot{h}_{+}-\left(1+h_+\right)\dot{h}_{\times}}{2\left(1-h_+^2-h_{\times}^2\right)},\;\;\;\quad\quad \Gamma_{13}^2=\frac{h_{\times}h'_{+}-\left(1+h_+\right)h'_{\times}}{2\left(1-h_+^2-h_{\times}^2\right)},\nonumber\\
\Gamma_{03}^3&=\frac{\left(1-h_+\right)\dot{h}_{+}-h_{\times}\dot{h}_{\times}}{2\left(1-h_+^2-h_{\times}^2\right)},\;\;\;\quad\quad \Gamma_{13}^3=\frac{\left(1-h_+\right)h'_{+}-h_{\times}h'_{\times}}{2\left(1-h_+^2-h_{\times}^2\right)},\nonumber\\
\Gamma_{22}^0&=-\frac{1}{2}\dot{h}_+,\;\qquad\qquad\qquad\qquad\Gamma_{22}^1=\frac{1}{2}h'_+,\nonumber\\
\Gamma_{23}^0&=-\frac{1}{2}\dot{h}_{\times},\;\qquad\qquad\qquad\qquad \Gamma_{23}^1=\frac{1}{2}h'_{\times},\nonumber\\
\Gamma_{33}^0&=\frac{1}{2}\dot{h}_{+},\quad\qquad\qquad\qquad\qquad\Gamma_{33}^1=-\frac{1}{2}h'_{+}.
\end{align}
Here, the prime denotes a partial derivative with respect to the coordinate $x$.\\
{\color{white}...}\\

\section{Constants $c_1$ and $c_2$}
The explicit expressions of the constants $c_1$ and $c_2$ in Eq.\,(\ref{probability}) are
\begin{align}\label{c1}
    c_1&=\left\{
    \left[1\!+\!\frac{E^4\!+\!2 E^2 (\alpha ^2\!-\!\beta ^2\!-\!1)}{(1+\alpha ^2+\beta ^2)^2}\right]^{\tfrac{1}{2}}\left[1\!+\!\frac{E^4\!+\!2 E^2 (\alpha ^2\!+\!\beta ^2\!-\!1)}{(1+\alpha ^2+\beta ^2)^2}\right]^{\frac{1}{2}}\right.\nonumber\\
    &\times\left[\frac{\beta\left(\alpha ^2\!+\!2 \alpha  \beta\!-\!\beta^2\right)}{2\sqrt{\alpha^2+1}}+\frac{\beta  \left(3\alpha^2\!+\!\beta^2\!+\!2\right)-2 \alpha  \left(\alpha ^2+1\right)}{2\sqrt{\alpha^2+1}(1+\alpha^2+\beta^2)}E^2\right]\nonumber\\
    &-\frac{4\beta^4+2\alpha\beta^3-4\alpha^2\beta^2}{(1+\alpha^2+\beta^2)^{\frac{3}{2}}}E^4-\frac{\beta^3(4\alpha^3-4\beta^3+6\alpha)}{(1+\alpha^2+\beta^2)^{\frac{5}{2}}}E^2\nonumber\\
    &+\frac{(2\alpha-\beta)(3\alpha+\alpha^3)+4}{(1\!+\!\alpha^2\!+\!\beta^2)^{\frac{3}{2}}}\!\nonumber\\
    &+\!\frac{\beta^4(2\!-\!\alpha\beta)\!+\!\alpha\beta(1\!-\!\alpha^2)^2\!-\!2\alpha^2(1\!+\!\alpha^2)\!-\!4}{(1+\alpha^2+\beta^2)^{\frac{5}{2}}}\nonumber\\
    &+\left[\frac{\alpha \left(\alpha ^2+2 \alpha  \beta-\beta ^2+2\right)+2 \beta}{(1+\alpha^2+\beta^2)^{\frac{3}{2}}}\beta E^3-\frac{\alpha ^2+2\alpha\beta-\beta^2}{\sqrt{1+\alpha^2+\beta^2}}\alpha\beta E\right]\nonumber\\
    &\times\left.
    \left[1\!+\!\frac{E^4\!+\!2 E^2 \left(\alpha ^2\!-\!\beta ^2\!-\!1\right)}{(1+\alpha^2+\beta^2)^2}\right]^{\frac{1}{2}}
    \right\}\left[1\!+\!\frac{E^4\!+\!2E^2(\alpha^2\!+\!\beta^2\!-\!1)}{(1+\alpha^2+\beta^2)^2}\right]^{-1}\!\!\!.
\end{align}
\begin{align}\label{c2}
    c_2&=\left\{
    -\left(\alpha^3+2\alpha^2\beta -\beta\alpha^2\right)\sqrt{\frac{\alpha ^2+\beta ^2+1}{4(1+\alpha^2)}}\right.\nonumber\\
    &\quad-\frac{\alpha \left(\alpha ^2+2 \alpha  \beta -\beta ^2+2\right)+2 \beta}{2\sqrt{1+\alpha^2}(1+\alpha^2+\beta^2)^{\frac{3}{2}}}E^4\nonumber\\
    &\quad-\frac{\alpha^2\beta\left(3 \alpha ^2+4\right)-\left(\alpha^5+\alpha ^3\right)+\alpha^2\beta ^3+2\alpha\beta ^2}{(\alpha ^2+1)(1+\alpha^2+\beta^2)^{\frac{3}{2}}} E^3 \nonumber\\
    &\quad+\frac{\alpha^3+2 \alpha ^2 \beta -\alpha  \beta ^2+\alpha +\beta}{\sqrt{(\alpha ^2+1)(1+\alpha^2+\beta^2)}} E^2\nonumber\\
    &\quad\left.-\frac{\alpha^2\beta\left(3\alpha ^2+2\right)-2\alpha\left(\alpha^2+1\right)^2+\alpha^2\beta ^3-2\alpha\beta ^2}{(\alpha ^2+1)\sqrt{1+\alpha^2+\beta^2}}E
    \right\}\nonumber\\
&\quad\times\left[1+\frac{E^4+2 E^2 \left(\alpha ^2-\beta ^2-1\right)}{\left(\alpha ^2+\beta ^2+1\right)^2}\right]^{-\frac{1}{2}}.
\end{align}

\end{document}